\documentclass[sigplan,screen]{acmart}

\usepackage[normalem]{ulem}

\usepackage{multirow}
\usepackage{xspace}
\usepackage{listings}
\usepackage{subcaption}
\usepackage{tikz}
\usepackage{paralist}
\usepackage{cleveref}
\usepackage{tcolorbox}
\usepackage[frozencache=true,cachedir=minted-cache]{minted}
\usepackage[subtle,lists,paragraphs,indent=normal,title=normal,bibbreaks=normal,
            floats=normal,mathspacing=normal,wordspacing=normal,
	    tracking=normal]{savetrees}

\setminted{fontsize=\scriptsize}

\newcommand\customtt[1]{{\small \texttt{#1}}}
\newcommand\grantno[1]{{\small #1}}
\newcommand\syscallno[1]{{\small #1}}

\newcommand{\tool}{Loupe\xspace}

\newcommand{\syscall}{system call\xspace}
\newcommand{\Syscall}{System call\xspace}
\newcommand{\syscalls}{system calls\xspace}
\newcommand{\SysCall}{System Call\xspace}
\newcommand{\SysCalls}{System Calls\xspace}
\newcommand{\Syscalls}{System calls\xspace}

\newcommand*{\eg}{e.g.,\@\xspace}
\newcommand*{\ie}{i.e.,\@\xspace}
\newcommand*{\vs}{vs.\@\xspace}
\newcommand*{\cf}{cf.\@\xspace}

\newcommand*\BC[1]{\tikz[baseline=(char.base)]{
	\node[shape=circle,draw,inner sep=0.15pt] (char) {\textcolor{black}{#1}};}}

\definecolor{mGreen}{rgb}{0,0.6,0}
\definecolor{mGray}{rgb}{0.5,0.5,0.5}
\definecolor{mPurple}{rgb}{0.58,0,0.82}
\definecolor{backgroundColour}{rgb}{0.97,0.97,0.97}

\newcommand{\todo}[1]{}
\renewcommand{\todo}[1]{{\color{red} TODO: {#1}}}

\definecolor{gray}{HTML}{ededed}
\colorlet{shadecolor}{gray}
\tcbset{
	colback=gray,
	boxrule=0.5pt,
	boxsep=1pt,
	left=3pt,
	right=3pt,
	top=3pt,
	bottom=3pt,
	sharp corners,
	colframe=black,
}

\newcommand{\insight}[1]{
	\begin{tcolorbox}
		\textbf{Insight:} #1
	\end{tcolorbox}
}

\copyrightyear{2024}
\acmYear{2024}
\setcopyright{rightsretained}
\acmConference[ASPLOS '24]{28th ACM International Conference on
Architectural Support for Programming Languages and Operating Systems,
Volume 1}{April 27-May 1, 2024}{La Jolla, CA, USA}
\acmBooktitle{28th ACM International Conference on Architectural Support for
Programming Languages and Operating Systems, Volume 1 (ASPLOS '24), April
27-May 1, 2024, La Jolla, CA, USA}
\acmDOI{10.1145/3617232.3624861}
\acmISBN{979-8-4007-0372-0/24/04}

\begin{document}

\title[\tool: Driving the Development of OS Compatibility Layers]{\tool: Driving the Development of\\OS Compatibility Layers}

\author{Hugo Lefeuvre}
\affiliation{
    \institution{The University of Manchester}
    \city{Manchester}
    \country{UK}
}

\author{Gaulthier Gain}
\affiliation{
    \institution{University of Liège}
    \city{Liège}
    \country{Belgium}
}

\author{Vlad-Andrei Bădoiu}
\affiliation{
    \institution{University Politehnica of Bucharest}
    \city{Bucharest}
    \country{Romania}
}

\author{Daniel Dinca}
\affiliation{
    \institution{University Politehnica of Bucharest}
    \city{Bucharest}
    \country{Romania}
}

\author{Vlad-Radu Schiller}
\affiliation{
    \institution{The University of Manchester}
    \city{Manchester}
    \country{UK}
}

\author{Costin Raiciu}
\affiliation{
    \institution{University Politehnica of Bucharest}
    \city{Bucharest}
    \country{Romania}
}

\author{Felipe Huici}
\affiliation{
    \institution{Unikraft.io}
    \city{Heidelberg}
    \country{Germany}
}

\author{Pierre Olivier}
\affiliation{
    \institution{The University of Manchester}
    \city{Manchester}
    \country{UK}
}

\begin{CCSXML}
<ccs2012>
<concept>
<concept_id>10011007.10010940.10010941.10010949</concept_id>
<concept_desc>Software and its engineering~Operating systems</concept_desc>
<concept_significance>500</concept_significance>
</concept>
</ccs2012>
\end{CCSXML}

\ccsdesc[500]{Software and its engineering~Operating systems}

\keywords{Operating Systems}

\renewcommand{\shortauthors}{Lefeuvre et al.}

\date{}

\begin{abstract}
Supporting mainstream applications is fundamental for a new OS to have impact.
It is generally achieved by developing a layer of compatibility allowing applications developed for a mainstream OS like Linux to run unmodified on the new OS.
Building such a layer, as we show, results in large engineering inefficiencies due to the lack of efficient methods to precisely measure the OS features required by a set of applications.

We propose \tool, a novel method based on dynamic analysis that determines the OS features that need to be implemented in a prototype OS to bring support for a target set of applications and workloads.
\tool guides and boosts OS developers as they build compatibility layers, prioritizing which features to implement in order to quickly support many applications as early as possible. We apply our methodology to 100+ applications and several OSes currently under development, demonstrating high engineering effort savings \vs existing approaches: for example, for the 62 applications supported by the OSv kernel, we show that using \tool, would have required implementing only 37 \syscalls \vs 92 for the non-systematic process followed by OSv developers.

We study our measurements and extract novel key insights. Overall, we show that the burden of building compatibility layers is significantly less than what previous works suggest: in some cases, only as few as 20\% of \syscalls reported by static analysis, and 50\% of those reported by naive dynamic analysis need an implementation for an application to successfully run standard benchmarks.
\end{abstract}

\maketitle

\section{Introduction}

An operating system is only as useful as the applications it can run.
Thus, developers of new OSes seeking to gather early performance numbers, to attract open source contributors, early investors, or to transition to real-world use~\cite{POSIX_PICOPROCESS, DRAWBRIDGE} need to provide support for existing applications.
Manually porting software~\cite{OSV_APPS, RUMP_PACKAGES} is only viable in the short term~\cite{HERMITUX, HERMITUX_TC}, hence developers of new and existing OSes must provide support for unmodified software by building compatibility layers~\cite{K42_LINUX, GRAPHENE, GRAPHENE_SGX, HERMITUX, HERMITUX_TC, UNIKRAFT, FUCHSIA, KERLA, GVISOR, OSV, ZEPHYR, LINUXULATOR, REACTOS, WSL, WINE, PROTON, WORKPLACE_OS, DOCKER_ON_ARM} that present applications with interfaces similar to that of popular OSes such as POSIX or the Linux kernel ABI.

Building a compatibility layer represents a non-negligible engineering effort~\cite{DRAWBRIDGE, POSIX_PICOPROCESS, K42_LINUX, OS_RISE_FALL, HERMITUX, HERMITUX_TC, UNIKRAFT, LUPINE} and involves 1) identifying the OS features (\syscalls, pseudo-files) required for a target application and 2) implementing these features.
This process is iteratively repeated for each application to support.
In this paper we focus on streamlining 1), the latter being generally OS-specific~\cite{K42_LINUX}.

We observe that, despite their cost, compatibility layers
are often developed in an ad-hoc fashion~\cite{K42_LINUX}: there is no systematic approach to determine and prioritize what OS features to develop and when, which applications to support, or to what extent used \syscalls should be implemented to achieve a desired degree of support.
This results in a significant amount of unnecessary engineering.

Past attempts at streamlining that process leverage static analysis~\cite{LINUX_API} and suffer from its drawbacks, heavily overestimating the set of OS features required to support an application.
For instance, while binary-level static analysis identifies that >100 \syscalls are required to conservatively support the entire superset of operations, configurations, and error handling code in Redis (much of which can be quite rarely used in practice, or simply irrelevant for an early prototype), we find that only 42 are actually needed to reliably pass its entire test suite, and just 20 to run \customtt{redis-benchmark}.

%Clearly, understanding and leveraging these intermediary levels of support is crucial to quickly and efficiently achieve application support.

Hence, OS designers often fall back to naive dynamic analysis, \eg using \customtt{strace}.
These tools fail to take into account common practices used in early OS development to save engineering effort: feature stubbing (returning \customtt{-ENOSYS}~\cite{ERRNO_ENOSYS} upon invocation, without implementing the feature), faking feature success (returning a success code without implementing the feature), and partial implementation of complex features~\cite{POSIX_PICOPROCESS, HERMITUX}.
Indeed, in early development, the goal is not to support every feature but rather core functionalities of target applications~\cite{K42_LINUX}.
For example, we find that more than half of the \syscalls invoked by Redis running the \customtt{redis-benchmark} can be stubbed or faked, and do not need to be implemented to support that application and workload.

We propose a systematic methodology based on dynamic analysis, centered around a novel tool called \emph{\tool}.
\tool measures, for an application and a given input workload (\eg a benchmark, test suite), which OS features really need to be implemented and which ones can be faked, stubbed, or partially implemented.
\tool also computes, given an OS under construction and a set of applications and workloads, an optimized development plan to support as many applications as possible with as little engineering effort as possible.

Dynamic analysis comes with its own challenges, in particular the difficulty to scale to numerous applications.
This is tackled by designing \tool to require as little effort as possible to integrate a new application, letting us present results for more than 100 applications in our evaluation.
Another challenge is how to detect OS features that can be stubbed, faked, or partially implemented.
This is addressed by leveraging Linux's \customtt{seccomp}~\cite{SECCOMP} and \customtt{ptrace}~\cite{PTRACE} tracing and interposition facilities to measure what OS features' implementation can be avoided with these techniques.

We run \tool on 100+ popular applications, and present examples of optimized Linux compatibility layer development plans obtained with \tool for three OSes under construction~\cite{FUCHSIA, KERLA, UNIKRAFT} with various levels of existing support for the Linux \syscall ABI.
We further measure the engineering effort savings obtained by using \tool to drive the development of compatibility layers.
Taking half the applications supported by OSv~\cite{OSV}, \tool reports that only 37 \syscalls are required to run them, \vs 92 for our estimation of the non-systematic process followed by OSv developers, and 142 for a process driven by \customtt{strace}-based dynamic analysis.

We study \tool's Linux API usage measurements for our set of applications.
This analysis brings many new insights.
We demonstrate that the minimal effort needed to provide compatibility is significantly lower than that determined by previous works using static analysis~\cite{LINUX_API}.
Our study shows that as much as 40-60\% of \syscalls found in application code do not need implementation to successfully run meaningful workloads, including full test suites.
We also find that many applications are resilient to stubbing, faking, and partial implementation of OS features.
We investigate the reasons behind it, and the impact of such practices on application performance and resource usage.
Finally, we study how the C library influences OS feature requirements.

In all, this paper makes the following contributions:
\begin{compactitem}

  \item A novel methodology to measure the minimum set of OS features that need implementation for a compatibility layer to support a set of applications and workloads, with the aim of minimizing development effort.

  \item \tool, a tool able to derive, for a given OS and target applications, an optimized OS feature support plan to run as many apps as possible, as early as possible.

  \item A demonstration of the engineering effort savings obtained with \tool, with examples of optimized feature implementation plans for 11 OSes under development.

  \item An analysis, using \tool, of the OS features required by a set of applications showing the lack of precision of past approaches and investigating common development practices in compatibility layer development.

\end{compactitem}

\tool is actively used in Unikraft~\cite{UNIKRAFT}, an open-source commercial OS, and has attracted the attention of several others.
Overall, this study brings a message of hope: contrary to what past work seems to suggest, a good degree of compatibility with existing applications can be achieved without immense engineering, provided we follow a focused and methodical approach.
\tool and our results are available online\footnote{\url{https://github.com/unikraft/loupe}\hfill/\hfill\url{https://github.com/unikraft/loupedb}} under an open-source license.

\section{Motivation and Approach}

\paragraph{Building Compatibility Layers for New OSes.}
Compatibility layers can be found in mature OSes for interoperability reasons~\cite{WSL, WINE, PROTON, LINUXULATOR, DOCKER_ON_ARM}, but also in a plethora of new/prototype/research OSes~\cite{K42_LINUX, GRAPHENE, GRAPHENE_SGX, HERMITUX, HERMITUX_TC, UNIKRAFT, FUCHSIA, KERLA, GVISOR, OSV, ZEPHYR, REACTOS, WORKPLACE_OS}.
Providing support for existing applications in these OSes is generally crucial~\cite{HERMITUX, HERMITUX_TC, UNIKRAFT, UNIKRAFT_LOGIN} to gather early performance numbers, to attract open source contributors, early investors, or transition to real-world use.
Manually porting software~\cite{OSV_APPS, RUMP_PACKAGES} is not sustainable in the long run, nor does it scale to a large amount of applications~\cite{HERMITUX, HERMITUX_TC}.
Hence, the developers of many new OSes resort to implementing compatibility layers.
Even considering OS models that choose to drop application compatibility for other gains (\eg performance), it is not uncommon
to see Linux versions of these models appear a few years after the seminal paper, with claims of stronger compatibility, \eg Popcorn Linux~\cite{POPCORN_ASPLOS} for the multikernel~\cite{BARRELFISH} or Graphene, Lupine and UKL~\cite{GRAPHENE, LUPINE, Raza2023} for the unikernel~\cite{MIRAGE_ASPLOS, CLICKOS}.

Building a compatibility layer is seen as a non-negligible engineering effort~\cite{DRAWBRIDGE, POSIX_PICOPROCESS, K42_LINUX, OS_RISE_FALL, HERMITUX, HERMITUX_TC, UNIKRAFT, LUPINE, LINUX_API}.
We investigated the compatibility layers present in several open-source OS projects~\cite{K42_LINUX, GRAPHENE, GRAPHENE_SGX, HERMITUX, HERMITUX_TC, UNIKRAFT, FUCHSIA, KERLA, GVISOR, OSV, ZEPHYR, REACTOS}.
Based on this study, and on our multiple years of experience providing Linux/POSIX compatibility in research OSes, we observe that compatibility layers are built in an ad-hoc, non-systematic (``organic'') way: developers select an application to support, determine the OS features it requires, and implement them~\cite{K42_LINUX}.
That process is repeated for each target application.
Because so many projects undergo the task of building compatibility layers~\cite{K42_LINUX, GRAPHENE, GRAPHENE_SGX, HERMITUX, HERMITUX_TC, UNIKRAFT, FUCHSIA, KERLA, GVISOR, OSV, ZEPHYR, REACTOS, WSL, WINE, PROTON, LINUXULATOR}, there is a need for tools to streamline that process.
The corresponding effort consists in 1) identifying OS features required by target applications and 2) implementing these features.
The latter task is known to be very specific to the new OS considered~\cite{K42_LINUX}, and can hardly be streamlined.
We show in this paper that the former task, identifying and prioritizing what OS features to implement, can be systematized and optimized.
Next, we motivate our method by explaining how past and current approaches are suboptimal.

% TODO if necessary, potential for trimming: "The corresponding effort consists in ..." is redundant with a sentence earlier.

\paragraph{Limitations of Static Analysis.}
Existing approaches measuring the usage of OS features by applications often rely on static analysis~\cite{LINUX_API, POSIX_API, HERMITUX, HERMITUX_TC, CONFINE, SYSFILTER, CHESTNUT, WAGNER_STATIC, ABHAYA, SAPHIRE}.
Static analysis is comprehensive: the set of features identified for an application includes all the ones that \emph{may} be invoked at runtime, under any possible workload, operation, or configuration, and traversing any possible error path.
Alas, static analysis is also conservative and yields many false positives: it overestimates OS features that will actually be invoked at runtime.

Static analysis can be performed on application sources or binaries.
Binary analysis~\cite{LINUX_API, POSIX_API, HERMITUX, HERMITUX_TC, SYSFILTER, CHESTNUT} scales well to a large number of applications because it targets a common format (\eg ELF binaries).
However, it suffers from a lack of precision due to the difficulty of extracting information from a binary~\cite{SYSFILTER}.
Such issues may be alleviated with source-level analysis~\cite{CONFINE, WAGNER_STATIC}, which is however not a panacea: it is language-specific, making it difficult to scale to many applications written in different languages.

Tsai et al.~\cite{LINUX_API} measure, using static binary analysis, the \syscall usage of the entire set of applications from an Ubuntu distribution.
The study concludes that to support 100\% of the distribution's packages, 272 \syscalls need to be implemented.
That number goes down to 81 \syscalls for the 10\% most popular applications.
These results suggest that a large implementation effort would be required for an OS aiming at supporting even a few applications.
As we demonstrate in the evaluation, both source- and binary-level approaches significantly overestimate the OS features required by an application to run popular workloads.
This is due to dead or unexecuted code, and the difficulty or impossibility to statically determine runtime-level information (\eg memory content such as function pointers).
Although all of these \syscalls would likely need to be implemented in a production-grade general-purpose OS, these numbers remain an upper bound of limited usefulness for OS designers in earlier development stages.

\paragraph{Limitations of Naive Dynamic Analysis.}
Dynamic analysis too has well-known drawbacks.
Its precision depends on the coverage of the input workload run during the analysis: if it is too low, some required OS features may not be identified.
It is also harder to fully automate, as there is a variable amount of manual effort required for each application to analyze (\eg selecting an input workload).
In this paper we refer to using a tool such as \customtt{strace}~\cite{STRACE} to trace OS features invoked by an application, as \emph{naive} dynamic analysis.
The main drawback of naive dynamic analysis is its failure to consider two techniques commonly used in early OS development~\cite{POSIX_PICOPROCESS}:
\begin{itemize}
\item Feature \emph{stubbing}: not implementing the feature and returning an error code (\customtt{-ENOSYS}: ``Not Implemented''~\cite{ERRNO_ENOSYS}) to the application when it invokes the feature.
  \item \emph{Faking} feature success: not implementing the feature and returning a success code (typically system-call specific) to the application upon invocation.
\end{itemize}

The two examples below are extracted from the source code of the HermiTux unikernel~\cite{HERMITUX}, where the \customtt{sigaltstack} \syscall is stubbed, and the \customtt{mprotect} \syscall is faked:

\begin{minted}[frame=single,framesep=5pt]{C}
long sys_sigaltstack(const stack_t *ss, stack_t *oss) {
  return -ENOSYS; // stubbed: not supported
}

long sys_mprotect(size_t addr, size_t len, uint64_t prot) {
  return 0; // faked: pretending success
}
\end{minted}

Many applications are resilient to the failure of OS features~\cite{POSIX_PICOPROCESS, Rinard2004} and will run correctly when stubbing and faking.
In this study, we show that many invoked OS features can avoid being implemented through these practices in the development stages of an OS.
This highlights the importance of faking and stubbing as an engineering practice: without it, showcasing a particular application use-case for a new OS concept would take significantly longer, or even be unattainable for a small-scale research project.
Despite of this, naive dynamic analysis does not typically consider stubbing and faking.
Naive dynamic analysis traces all features and sub-features invoked by an application, independently of the fact that they can be stubbed/faked or not for a given workload.
Thus, OS designers typically rely on trial and error to determine which features they need to implement first, and which ones they can fake or stub.

\paragraph{When to Stub or Fake and When not To?}

The reliance on stubbing and faking as a development practice in transitional OS development stages introduces a pivotal question: \emph{when to stub and fake, and when not to?}
This question is driven by two sources of concern:

\begin{itemize}
\item Impacting stability.
Although guaranteed stability of entire applications is not a primary goal in the early development stages of an OS, faking and stubbing must not impact the stability of relevant application features.
Failing to do so would negate the benefits of faking and stubbing by creating an additional debugging cost.

\item Impacting performance metrics.
Early OS prototypes must be comparable to full-fledged mainstream OSes; this is especially true for research OSes.
Impacting performance metrics by faking or stubbing would defeat the purpose of the OS prototype by making it impossible to fairly evaluate its performance advantage or cost.
For instance, stubbing or faking an expensive and relevant security feature may provide an unfair advantage to an early OS prototype \vs a full-fledged OS that implements it.
\end{itemize}

Non-systematic, trial-and-error-based approaches are especially prone to fall into stability and performance pitfalls.
Although important, these concerns have been little discussed by works which rely or relied on faking and stubbing.

\paragraph{Breaking the Status Quo with \tool.}
We aim to propose a systematic and adaptive method to determine which OS features to implement first.
Our goal is to help OS designers transition from \emph{no support} towards \emph{full support} to run as many applications as early as possible.

Overall, dynamic analysis is better suited to the problem we aim to solve, being able to evaluate the concrete impact of both stubbing and faking, and providing fine-grain, per-workload results.
The coverage issue of dynamic analysis is a nonproblem in our context: in early development phases of an OS, the goal is not to support every feature but rather core functionalities of target applications~\cite{K42_LINUX}, which are easily exercised by standard benchmarks and test suites.
Dynamic analysis is precisely suited because it is adaptive: as support progresses, workloads can be extended to cover more and more application features.
We are left with two challenges: the difficulty to 1) scale to numerous applications, and to 2) detect OS features that can be stubbed, faked, or partially implemented without impacting stability or performance for relevant application features.
As we detail in \S\ref{sec:tool}, we solve the former by designing \tool to require as little effort as possible to integrate a new application (at most writing a Dockerfile and test script), letting us present results for 100+ applications in \S\ref{sec:syscall-analysis}.
We address the latter challenge by leveraging Linux's \customtt{seccomp}~\cite{SECCOMP} and \customtt{ptrace}~\cite{PTRACE} facilities to measure what OS features can be stubbed, faked, or partially implemented for a given application workload.
To maintain stability and performance metrics for the evaluated workload, \tool replicates the analysis several times in containerized environments, and offers a framework for identifying performance regressions on various generic and application-specific metrics.
We further discuss stability in \Cref{subsec:resilience-stubfake,sec:discussion}, and performance impact in \Cref{subsec:fakestub-impact}.

\section{\tool: Accurate Run-time Analysis of OS Feature Usage}
\label{sec:tool}

To accurately quantify what OS features are needed to gradually support a set of popular Linux applications, and to measure to what extent static and naive dynamic analysis overestimate these requirements, we built \emph{\tool}.
Loupe is a dynamic analysis tool that hooks into each OS feature used by applications at runtime, analyzing the application's behavior as it simulates different degrees of compatibility.
Unlike existing naive dynamic analysis tools (\eg { }\customtt{strace}), \tool is built as a framework specifically meant for OS feature support analysis.
It supports identifying what \syscalls and pseudo-files are used by a given application, and determining which can be faked, stubbed, or partially implemented.
\tool focuses on reliable and reproducible results, and supports easy integration into existing build systems and complex test suite systems.
Finally, \tool can process measurement data for a set of applications and output targeted OS feature support plans.
We implemented a prototype of \tool in 2.5K LoC of Python, and 500 LoC of C (used for \customtt{seccomp} and \customtt{ptrace} hooks).
In this section, we summarize the functioning and architecture of \tool (\S\ref{subsec:loupe-overview}), detail our approach to evaluate the success and performance of application runs (\S\ref{subsec:testapps}), and conclude with details on various aspects of \tool (\S\ref{subsec:loupe-detail}).

\subsection{\tool Overview}
\label{subsec:loupe-overview}

\begin{figure}
\center
\includegraphics[width=\linewidth]{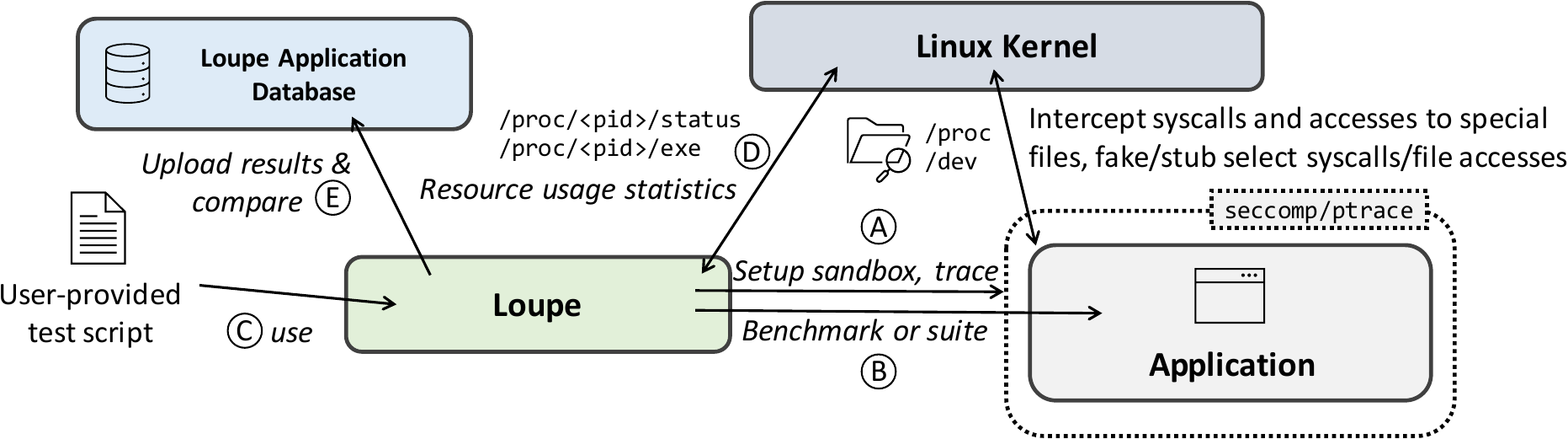}
\caption{
\tool architecture diagram.
}
\label{fig:architecture}
\end{figure}

Using \tool to measure the OS feature usage of a new application is straightforward.
Users provide \tool with 1) the application binary whose OS feature usage needs to be measured, and 2) a per-application \emph{test script}, responsible for providing external input to the application, and measuring the performance and success of each run.
\tool operates on binaries, so there are no language or compiler restrictions on either the application or the test script.
Provided this, \tool evaluates the OS feature usage of the application feature by feature.
Each used \syscall and pseudo-file is tested for one of two modes in separate runs: 1) \emph{stubbing} the \syscall, \ie do not run the \syscall and return \customtt{-ENOSYS} or 2) \emph{faking} it, \ie return a success code without running the \syscall.
Once all OS features have been tested, a final run confirms that the analysis performed on a per-feature basis holds when all features are considered.
In the event of a failure, users can use \tool to alter subsets of \syscalls to find the culprits, a process which could be automated in future works.

We now detail the behavior of \tool \emph{for each run}, as visualized in \Cref{fig:architecture}.
\tool first simultaneously sets up tracing and sandboxing (\BC{A} on Figure~\ref{fig:architecture}) and starts the application (\BC{B}) using the \customtt{seccomp}~\cite{SECCOMP} and \customtt{ptrace}~\cite{PTRACE} Linux tracing and interposition facilities.
Once the application has been started, \tool uses the test script to feed the application with inputs (\eg generating client requests for a server application) and gather performance numbers (\BC{C}), all the while recording data on resource usage via \customtt{/proc} (\BC{D}).
Using the hooks set up in \BC{B}, \tool intercepts each \syscall invoked by the application, and tests it for one of the two previously described modes.
At the end of the run, \tool determines the success of the application using the return code of the test script (more in \S\ref{subsec:testapps}).
Accesses to pseudo-files are hooked and disabled, stubbed, or faked similarly by catching \syscalls from the \customtt{open} family (see \S\ref{loupe-approach-pseudofiles}).

In order to maximize the reliability and reproducibility of the results, each analysis is performed multiple times in containerized replicas, and the result of the analysis is conservatively updated to take all results into account.
The number of replicas (\emph{3} by default) and whether they run in parallel (\emph{no} by default) can be configured to suit different applications, accuracy needs, and available hardware.

Finally, OS developers can specify the \syscalls supported by their OS in CSV form, and \tool will recommend which OS features to implement, stub, or fake, to support a set of applications selected among those measured by the tool.
\tool will prioritize the list of features to indicate which should be implemented first in order to support as many applications, as early as possible.
\tool's measurements can optionally be shared in an online database (\BC{E}).

\subsection{Evaluating Success and Performance}
\label{subsec:testapps}

\tool builds on the premise that users are able to describe a workload that they want to support for a given application.
Loupe then tells the user which precise set of system calls they have to support (and how) to be able to run that workload reliably, i.e., over multiple runs without observable functional and non-functional issues.

\paragraph{Describing Workloads.}
Workloads describe the feature set that must be supported in a given application.
\tool users express workloads in \emph{test scripts}, responsible for supplying external input, if required by the application, and detecting the success of a run.\footnote{
Some programs do not require input and determine success by themselves or via a wrapper script (\eg test suites).
If so, the test script is \emph{practically included} in the application and need not be passed separately.
\tool supports this.
Since this is similar to the general case, we do not further discuss it here.}
Test scripts may materialize any type of workload: simple health checks (\eg for a web server: can the application process a simple HTTP query?), benchmarks, test suites, or even fuzzing.
If specific error cases or application features must be supported, then the test script must also exercise them as part of the run.
In this paper we explore health checks, benchmarks, and test suites.
Each workload may be relevant at different stages in the development life cycle of a new OS.
Workloads correspond to different levels of guarantee of application stability; they can be evolved as support progresses, until complete compatibility can be provided to ensure stable application behavior in all circumstances.

\paragraph{Defining ``Success''.} \label{par:gather-nonfunctional-metrics}
A run is considered successful when the application terminates and the test script exit code indicates success.
Crashes, or unresponsiveness are considered as generic failure signs.
The notion of generic failure can be extended to unusual resource usage, or even unusual filesystem or network usage, which \tool can observe without understanding application semantics.
Generally however, the notion of success or failure is application-specific and inseparable from the workload itself: \eg outputs on the standard output/error channels or logs that do not correspond to normal application behavior, or altered performance (\eg throughput, latency, packet loss rate).
Application-specific success criteria must be evaluated by the test script.
An example of a test script for Nginx benchmarked with \customtt{wrk} is shown below:

\begin{minted}[frame=single,framesep=4pt]{bash}
#!/bin/bash
# [...] omitted helpers (including is_failed and grep_req_per_sec)

b=$(wrk http://localhost:8080 -d10s | grep_req_per_sec)
if [[ $(is_failed $b $?) ]]; then exit 1;
else echo $b; fi
\end{minted}

\customtt{is\_failed()} is responsible for detecting failures, left out above for space reasons.
When performing a simple health check, the function verifies that the throughput is non-zero.

We implemented detection of unusual resource usage and performance in our prototype.
\tool records application resource usage (maximum resident size and open file descriptors) via \customtt{/proc} and compares results over multiple runs when stubbing or faking.
Similarly, when performing a performance benchmark, test scripts return the relevant performance number (which can be any application-specific performance metric), and \tool ensures that the performance does not incur a statistically significant variation from the full-fledged baseline.
Together, resource usage and performance checks can provide insights into the impact of stubbing or faking features, and particularly increase the confidence on the correctness (or incorrectness) of faking and stubbing.
We further discuss performance and resource usage in \S\ref{subsec:fakestub-impact}.

\subsection{\tool in Detail}
\label{subsec:loupe-detail}

We now discuss various aspects of \tool that are relevant in this paper: supporting vectored \syscalls and pseudo-files, making \tool easy to use in many applications, how long \tool analyzes take, and sharing analysis results.

\paragraph{Vectored \SysCalls.}
Identifying OS features at the granularity of an entire \syscall is sometimes too coarse, considering vectored \syscalls (\eg { }\customtt{ioctl}, \customtt{fnctl}) and \syscalls with several functionalities that may be partially implemented in a compatibility layer (\eg { }\customtt{mmap}, or \customtt{madvise}).
In such cases, \tool can also disable, stub, and fake \syscalls based on \textit{individual \syscall parameters}, allowing users to easily explore partial implementations at a fine granularity.
The output is a list of \syscalls along with their used sub-features, and whether they can be faked or stubbed.

% TODO link to a pseudo file manpage would be good here
\paragraph{Pseudo Files.} \label{loupe-approach-pseudofiles}
Part of the Linux API is offered through pseudo-files such as \customtt{/dev/random}.
\tool is able to detect usage of such special files by pattern matching the arguments of certain \syscalls (\eg { }\customtt{open}, \customtt{openat}) against paths (\eg { }\customtt{/dev}, \customtt{/proc}).
\tool can also fake or stub \syscalls accessing these files, enabling users to track which special files require an implementation for applications to run.

\paragraph{Testing Framework Integration.}
Dynamic analysis tools can be difficult to integrate in application testing frameworks.
Test suites, for instance, may start the application multiple times, from complex scripts, from different call points~\cite{SQLITE_SUITE, REDIS_SUITE}.
Calling a naive analysis tool like \customtt{strace} requires manual changes, along with additional logic to gather and merge results obtained from the multiple runs triggered by the test suite.
Calling the tool on the test suite itself (\eg { }\customtt{strace make test}) is not effective either, as the test suite may call external tools whose OS feature usage is not part of the application's.
For instance, the Ruby test suite makes extensive calls to \customtt{git} to set up test environments; the OS feature requirements of \customtt{git} should not be included into the application's.
We tackle this problem with a whitelist system: when run on a wrapper (\eg a test suite), users can specify which binaries are that of the application and should be considered in the analysis.
\tool then tracks all children processes, checking the binary path upon \customtt{exec}, to ignore any \syscall originating from a binary that does not correspond to the specified one(s).
This allows, for instance, unmodified analysis of test suites run via \customtt{make test}; \tool simply executes the Makefile and only considers \syscalls executed by the appropriate binary.

\paragraph{Debhelper Integration.}

To further simplify running \tool on many applications, we integrated \tool into the Debhelper~\cite{DEBHELPER} Debian package build system.
\tool can build Debian packages and run on the package's \customtt{dh\_auto\_test}~\cite{DH_AUTO_TEST} rule which, if provided by the package, executes the target application's test suite.
Combined with the previous technique, which \tool can leverage by listing the package's binaries, we can significantly reduce the cost of testing applications.
Running \tool on the Lighttpd, Memcached, and webfsd test suites, for instance, is fully automated this way.

\paragraph{\tool Run Time.}

The runtime of a full \tool analysis is $(2+(2*t*s))*\lceil\frac{r}{p}\rceil$ with: $t$ the application workload runtime, $s$ the number of distinct \syscalls (and pseudo-files, if enabled in the analysis) executed by the application under the specific workload, $r$ the number of replicas, and $p$ the number of replicas executed in parallel.
$2+$ corresponds to the initial run to discover executed \syscalls, and to the final run to confirm the analysis.
$2*$ corresponds to the ``stubbing'' and to the ``faking'' run for each \syscall.
The overall runtime is therefore dominated by the length and complexity of the application workload; it varies from about 4~minutes for a fast Nginx health check, to 50~minutes for the Lighttpd test suite, and 1-1.5~days for the SQLite test suite (by far the largest we encountered, running \emph{millions} of tests~\cite{SQLITE_SUITE}).
These run times are reasonable: porting cost for a single application often reach multiple weeks or months in early OS development stages~\cite{UNIKRAFT} and, as we expand next, this is a one-time cost.

\paragraph{Sharing \tool Results.}
\label{subsec:ci}
Thanks to the techniques described previously, \tool test scripts are easy to write; 2-30 minutes on average according to the expertise of the user, most of it spent on understanding how to run and test the application.
The main barrier to running \tool on a large number of applications is runtime.
Nevertheless, as we described previously, the results are final for a fixed build of the software, its workload, dependencies, kernel, and test script.
To leverage this, we have set up a shared online database that can be populated and looked up by any individual running \tool or interested in its results.
\tool can automatically submit results to the database along with metadata (\BC{E} in Figure~\ref{fig:architecture}).
We envision that in the long run, this database will contain results for a wide range of applications, helping OS and application developers to study OS features usage patterns, build compatibility layers, and more, without even the runtime cost mentioned previously.

\section{\tool: OS Feature Support Guide}
\label{subsec:support-plan}

For space reasons, we set aside pseudo files and focus on \syscall support, as it represents the majority of the engineering effort in building compatibility layers~\cite{HERMITUX, UNIKRAFT, KERLA}.

\subsection{Examples of Support Plans}

We ran \tool on a total of 116 applications with various workloads including standard benchmarks (\eg \texttt{wrk} for web applications, \texttt{redis-benchmark}).
We choose a selection of representative applications from OpenBenchmarking.org~\cite{OPENBENCHMARKING}, as well as various other sources~\cite{OSV_APPS, RUMP_PACKAGES, UNIKRAFT_APPS}\footnote{Our artifact includes a list of all applications and support plans for 11 OSes: \url{https://github.com/unikraft/loupedb/blob/staging/ASPLOS24-supp.pdf}}.
Leveraging these measurements, \tool guides the process of developing a compatibility layer by giving a prioritized list of \syscalls to implement/stub/fake.
Specifically, given (1) the state of a partially Linux-compatible OS in terms of \syscalls supported (a simple text file with one line per supported \syscall) and (2) a set of target applications to support, \tool can output an incremental support plan listing the order in which missing \syscalls should be implemented/faked/stubbed in order to enable compatibility with a maximum of applications as early as possible.

\begin{table}
\caption{Step-by-step support plans for 3 OSes.}
\setlength{\tabcolsep}{3.5pt}
\label{tab:unikraft-plan}
{\footnotesize
\begin{tabular}{|c|p{1.9cm}|p{1.86cm}|p{1.37cm}|p{1.6cm}|}
\hline
Step & Implement & Stub & Fake & Support for\dots\\
\hline
\hline
%0 & - & -\\
%\hline
\multicolumn{5}{|c|}{\textbf{Unikraft} (commit \texttt{7d6707f}, supports 174 syscalls)}\\
\hline
0 & - & - & - & (12 apps)\\
\hline
1 & 290 & 273, 218, 230 & - & + Memcached\\
\hline
2 & 218 & - & - & + H2O\\
\hline
3 & 283, 27 & 186 & - & + MongoDB\\
\hline
\hline
\multicolumn{5}{|c|}{\textbf{Fuchsia} (commit \texttt{5d20758}, supports 152 syscalls)}\\
\hline
0 & - & - & - & (10 apps)\\
\hline
1 & 33 & 273, 302, 105 & - & + Lighttpd\\
\hline
2 & 302 & 230 & - & + Memcached\\
\hline
3 & - & 99, 222, 223 & - & + HAProxy\\
\hline
4 & 105 & 40 & 128, 99, 27 & + Nginx\\
\hline
5 & 128, 99, 27 & - & - & + MongoDB\\
\hline
\hline
\multicolumn{5}{|c|}{\textbf{Kerla} (commit \texttt{73a1873}, supports 58 syscalls)}\\
\hline
0 & - & - & - & (4 apps)\\
\hline
1 & 56, 257, 54 & (17 syscalls) & 47 & + Httpd\\
\hline
2 & 10 & - & 302 & + Weborf\\
\hline
3 & 8, 21, 87 & - & 25 & + SQLite\\
\hline
4 & 232, 233, 302 & (9 syscalls) & 288, 213 & + HAProxy\\
\hline
5 & 17, 213, 262 & 95 & - & + Redis\\
\hline
6 & 291 & 105, 106, 116, 293 & - & + Lighttpd\\
\hline
7 & 288, 290 & 32 & 102 & + H2O\\
\hline
8 & 46 & 230 & - & + Memcached\\
\hline
9 & 105, 18, 53, 106 & 40 & 92, 130, 107, 273, 116, 157 & + Nginx\\
\hline
10 & 104, 107, 108, 102 & - & - & + Webfsd\\
\hline
11 & 128, 99, 229, 27, 73, 202, 283 & 131 & 137 & + MongoDB\\
\hline
\end{tabular}
}
\end{table}

We enabled \tool to generate support plans for all 116 applications we measured, for 11 OSes under development: Unikraft~\cite{UNIKRAFT}, Google Fuchsia~\cite{FUCHSIA} and Zephyr~\cite{ZEPHYR}, Kerla~\cite{KERLA}, HermiTux~\cite{HERMITUX}, Google gVisor~\cite{GVISOR}, Graphene/Gramine~\cite{GRAPHENE, GRAMINE_WEBSITE}, FreeBSD Linuxulator~\cite{LINUXULATOR}, Browsix~\cite{BROWSIX}, OSv~\cite{OSV}, and Linux nolibc~\cite{NOLIBC}.
To illustrate this functionality, we present here a subset of these results (for space reasons): we consider recent versions of 3 OSes: Unikraft, Fuchsia and Kerla, and a target set of 15 popular cloud applications.
The support plans are presented in Table~\ref{tab:unikraft-plan}.
The number of steps to reach support for all 15 apps is directly linked to the maturity of the OS: Unikraft for example has initial support for 12 applications and
requires only 3 steps to reach full support, while Kerla, with initial support for only 4 applications, requires 11 steps.
\tool's incremental support plans optimize the development of compatibility layers by breaking down the effort into small steps (>80\% of which requiring to implement 1-3 \syscalls), unlocking support for an application after each step.
The support plans in Table~\ref{tab:unikraft-plan} target a small set of applications for space reasons.
Full support plans for each of the 11 OSes we target, for all 116 applications in our database, are larger: 35 steps for Fuchsia, 32 for Unikraft, and 79 for Kerla.

\subsection{Engineering Effort Savings}

To estimate the engineering effort savings that an OS project would enjoy while building a compatibility layer with \tool rather than in an ad-hoc, organic fashion, we designed the following experiment:
we select a large set of 62 applications supported by a popular experimental OS, OSv~\cite{OSV}, from the OSv-Apps repository~\cite{OSV_APPS}.
We then estimate the order in which these applications were organically supported by the OS.
For that we use git metadata to track the creation date of the folder corresponding to each app in the repository.
We then derive from the order in which applications were supported, the organic order in which \syscalls had to be implemented by OSv developers.
Because stubbing/faking OS features are well-known practices~\cite{POSIX_PICOPROCESS}, and because there are traces of their usage in OSv's codebase~\cite{OSV_STUB}, we assume that OSv developers used stubbing and faking as much as possible.
We can then derive, in chronological order, the number of \syscalls that were implemented by OSv developers, and the evolution of the number of supported applications.
We also compute these numbers for a hypothetical optimized compatibility layer development process that would be guided by \tool's support plan, which would also take stubbing/faking into account, as well as a naive approach that would implement every \syscall traced by dynamic analysis, without stubbing/faking.

\begin{figure}
    \center
    \includegraphics[width=.45\textwidth]{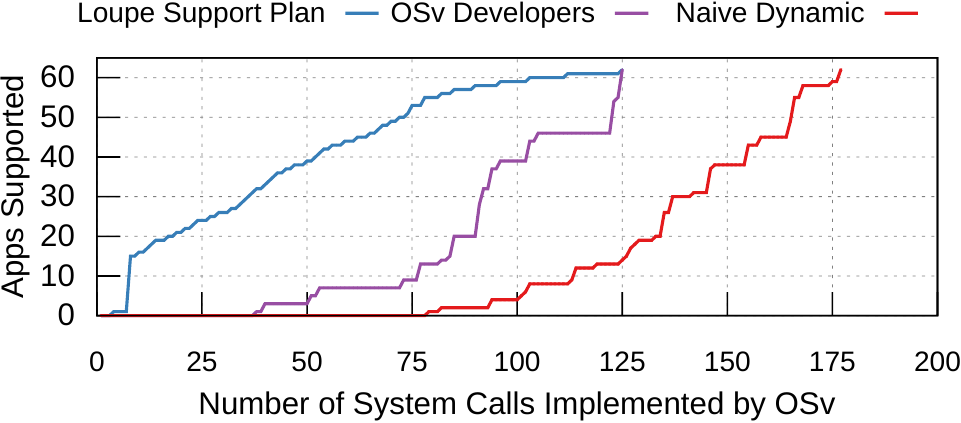}
    \caption{Evolution of the number of applications and \syscalls supported by OSv assuming 1) a support plan generated with \tool, 2) organic development based on git history, 3) measurement with naive dynamic analysis without stubbing/faking. Higher values indicate more applications supported for the same effort.}
    \label{fig:vlad-graph}
\end{figure}

These results are presented on \Cref{fig:vlad-graph}.
As one can observe, \tool would have heavily optimized the process of implementing OSv's support for the target application set, leading to more applications supported earlier and with less engineering effort \vs our estimation of the organic process undertaken by OSv's developers.
For example, to support half (31) of the applications, with \tool only 37 \syscalls need to be implemented, \vs 92 for the organic process.
The naive method relying on dynamic analysis without stubbing/faking requires even more engineering effort: to reach 31 applications, 142 \syscalls would need to be implemented.

Our method to estimate engineering efforts makes a few simplifications.
The real order in which applications were supported by OSv is likely not exactly that of folder creation in the OSv-Apps repository.
We repeated the study using the date of the \emph{last commit} in each application's folder to determine the order; results were similar.
The effort to implement \syscalls is also variable according to which \syscall is targeted: the x-axis in \Cref{fig:vlad-graph} is non-uniform since not all \syscalls have the same implementation cost.
However, we believe these results provide a sufficiently solid estimation of the engineering effort reduction that \tool can bring to demonstrate its usefulness.

\section{Analyzing the Linux API with \tool}
\label{sec:syscall-analysis}

Here we study the Linux API usage results obtained using \tool for the 116 applications considered in our study.
We aim to answer the following research questions:

\begin{itemize}
    \item How important is the accuracy gap between \tool's method vs naive dynamic analysis (\customtt{strace}) and static analysis?

    \item When building a Linux compatibility layer, which \syscalls must be implemented, and which ones can be commonly faked or stubbed?
        What is the absolute minimum set of \syscalls that must be implemented for a test suite to correctly run?

    \item What are the most important \syscalls, \ie the ones whose implementation is required by most applications?

    \item Why can some \syscalls be faked or stubbed?
        Does it impact performance or resource usage metrics?

\item How much do the \syscall requirements of applications and standard libraries evolve over time?
\end{itemize}

For space reasons, we concentrate on \syscalls and set aside results regarding special files and vectored \syscalls.

\subsection{Analysis Method: Static \vs Dynamic}

\paragraph{\tool \vs Naive Dynamic Analysis.}
We computed the \emph{API importance} of each \syscall as reported by \tool and by naive dynamic analysis.
API importance~\cite{LINUX_API} represents the probability that in our 116 applications data set, a \syscall is required by at least one application in that set.
A \syscall is defined as required for an application if it is traced with dynamic analysis, and if it is traced and can neither be stubbed nor faked with \tool.

\begin{figure}
  \center
  \includegraphics[width=0.9\linewidth]{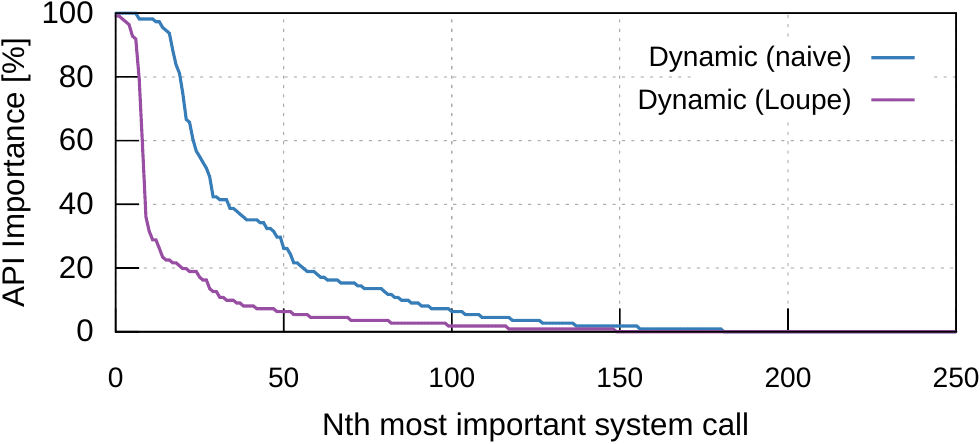}
  \caption{API importance for dynamic analysis with \tool and a naive approach (= no stubbing/faking).}
  \label{fig:tsai-like-full}
\end{figure}

\Cref{fig:tsai-like-full} visualizes our results.
They show that naive dynamic analysis severely overestimates the amount of \syscalls required to support applications.
\tool reports a total of 148 \syscalls requiring implementation to support 100\% of our 116 applications, \vs 180 \syscalls for a naive analysis.
The 25 most commonly required \syscalls are present in more than 80\% of the applications with \tool, and in less than 50\% with naive dynamic analysis.

\paragraph{\tool \vs Static Analysis.}
\label{sec:vs-static}

% Placed here because of suboptimal latex placement...
\begin{figure*}
  \center
  \includegraphics[width=0.9\textwidth]{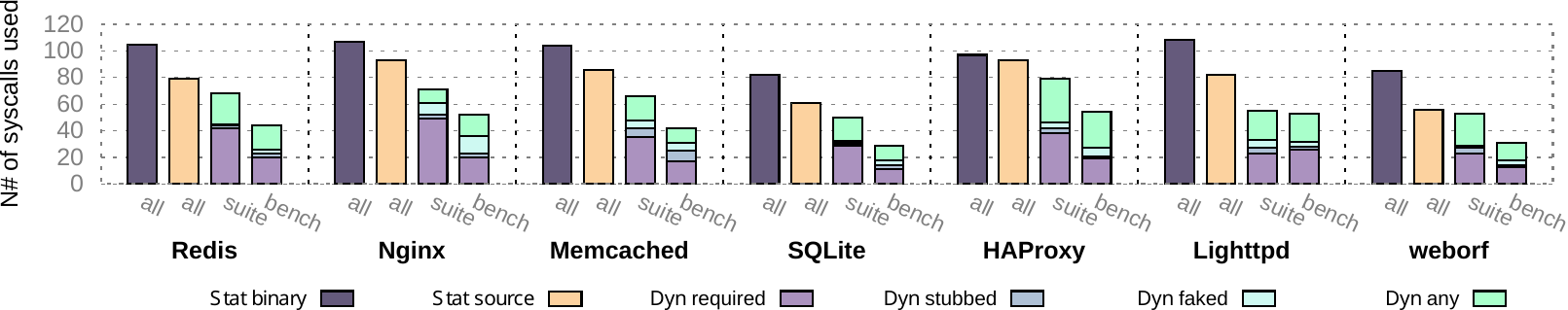}
  \caption{Number of \syscalls statically identified and dynamically traced by \tool for applications running standard benchmarks (\emph{bench}) and test-suites (\emph{suite}). Traced \syscalls are broken down into those that can be stubbed, faked, either faked or stubbed (\emph{any}), and those that can neither be faked nor stubbed (\emph{required}).
}
  \label{fig:syscall-support}
\end{figure*}

\begin{figure*}
  \centering
  \begin{subfigure}{.485\textwidth}
    \centering
    \includegraphics[width=\linewidth]{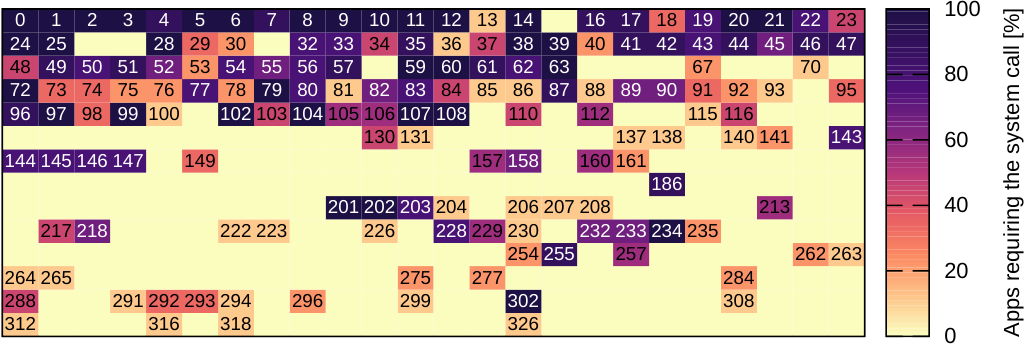}
    \caption{Static analysis, binary level.}
    \label{fig:heatmapstaticbinary}
  \end{subfigure}%
  \begin{subfigure}{.485\textwidth}
    \centering
    \includegraphics[width=\linewidth]{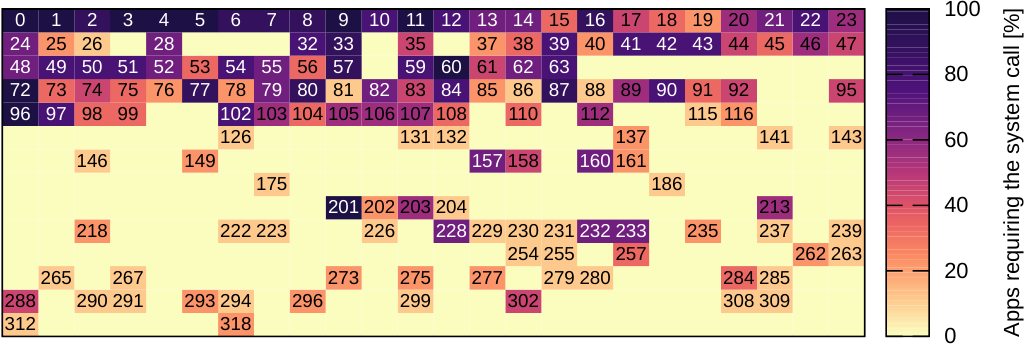}
    \caption{Static analysis, source level.}
    \label{fig:heatmapstaticsource}
  \end{subfigure}%
  \vskip\baselineskip
  \begin{subfigure}{.485\textwidth}
    \centering
    \includegraphics[width=\linewidth]{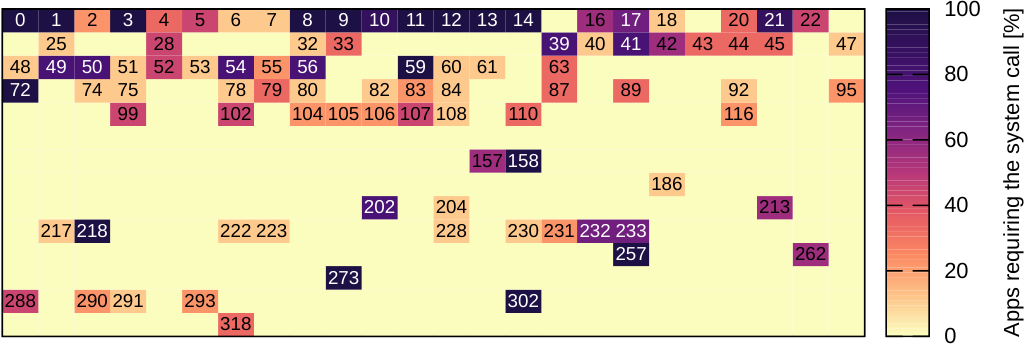}
    \caption{Dynamic analysis, traced.}
    \label{fig:heatmapdynexecuted}
  \end{subfigure}
  \begin{subfigure}{.485\textwidth}
    \centering
    \includegraphics[width=\linewidth]{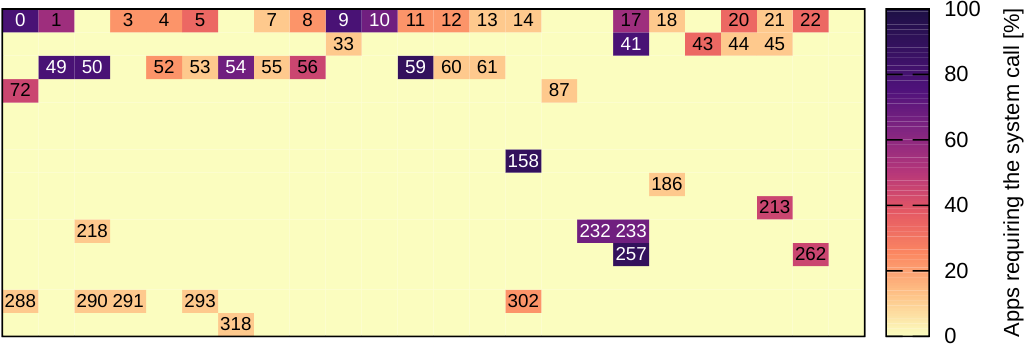}
    \caption{\tool's dynamic analysis, required (excluding stubbed/faked).}
    \label{fig:heatmapdynrequired}
  \end{subfigure}
  \caption{
  \Syscalls identified by static binary, static source, naive dynamic \textit{traced} (all \syscalls detected), and \tool's dynamic \textit{required} (those that cannot be stubbed/faked).
  Each box represents a Linux \syscall and its number.
  }
  \label{fig:heatmaps}
\end{figure*}

We faced scalability issues when trying to apply binary- and source-level static analysis tools to our large data set of 116 applications.
There exists no source-level tool able to identify \syscalls for all the relevant programming languages.
We also attempted to run several binary-level tools and experienced a high level of failures (close to 50\%) skewing the results.
Hence, we fall back on selecting a subset of applications from our data set for comparison between static analysis and \tool.

We select 7 popular cloud applications that support standard benchmarks and ship with comprehensive test suites: Redis, Nginx, Memcached, SQLite, HAProxy, Lighttpd, and Weborf.
To gather results for static analysis we use the source- and binary-level tools made available by Unikraft~\cite{STATIC_ANALYSIS_SRC, STATIC_ANALYSIS_BIN}.
Figure~\ref{fig:syscall-support} details the amount of \syscalls identified in each application by each method.
Both static analysis techniques severely overestimate the number of \syscalls actually needed to run the benchmarks and test suites.
The minimum number of \syscalls identified by \tool as required for these applications varies around 20 for benchmarks, and 20-40 for test suites.
Both static binary and source analysis methods report numbers that are generally between 5x and 2x higher.
For example, on Redis, binary-level static analysis identifies 103 \syscalls \vs 68 dynamically traced ones from the test suite, and \tool further indicates that more than a third of these can be stubbed/faked.
This observation can be generalized to all other applications.
Overall these results show that the effort to provide comprehensive support of core features and even full test suites is much lower than suggested by previous work based on static analysis~\cite{LINUX_API}.

Figure~\ref{fig:heatmaps} details which \syscalls are detected by the various analysis techniques when applied to the 7 applications running benchmarks.
Once again the overestimation of static and naive dynamic analysis is clear, compared to the results obtained with \tool.
Regarding static analysis, operating on the binary only yields more \syscalls compared to targeting the sources.
Concerning dynamic analysis, a non-negligible amount of \syscalls can be stubbed/faked, confirming the benefits of \tool \vs naive dynamic analysis.
We investigate faking/stubbing more in details next.

\insight{
  Static and naive dynamic analysis both highly overestimate the engineering effort needed to build a compatibility layer for a target set of applications.
}

\subsection{Resilience to Stubbing and Faking}
\label{subsec:resilience-stubfake}

As visualized in Figure~\ref{fig:syscall-support}, we find that, on average, the proportion of invoked \syscalls that can be stubbed or faked is 46\% for test suites (ranging from 31\% for Nginx to 58\% for Lighttpd), and 60\% for benchmarks (from 51\% for Lighttpd to 65\% for HAProxy).
This shows that the effort required to provide strong support of core features (\ie those covered by test suites) for these popular applications is certainly lower than suggested by previous work, and is even lower when considering support for benchmarks only (needed for evaluation in research papers).
The difference between Figure~\ref{fig:heatmapdynexecuted} and ~\ref{fig:heatmapdynrequired} clarifies this, highlighting which \syscalls can commonly be stubbed and faked.
We observe broadly two categories:

\begin{itemize}
\item \textbf{Low range \syscalls (\syscall ID \textasciitilde\emph{<~150})}, representing the majority of \syscalls detected by all analysis methods. This is unsurprising as these \syscalls correspond to core services that have been present in the Linux feature set for a long time, such as basic network \syscalls (\customtt{bind}, \customtt{accept}, etc.).
\item \textbf{Higher range \syscalls (ID \textasciitilde\emph{>~150})}, where a small set of  popular \syscalls are invoked corresponding to more modern but prominent functionality concerning multithreading (\customtt{futex} -- 202, etc.), scalable I/O (\customtt{epoll} family -- 213, 232, 233), as well as new variants of core \syscalls (\customtt{openat} -- 257, \customtt{prlimit64} -- 302, etc.)
\end{itemize}

Though \syscalls from both categories can be stubbed or faked, \syscalls with higher numbers are better candidates: out of the lower half of used \syscalls (46 \syscalls with number \emph{<~63}), 13 \syscalls can always be stubbed \vs 30 for the upper half (46 \syscalls with number \emph{>~63}).
This is because these map to more recent, generally less critical functionalities; we expand on this next.

\insight{
  Though applications may invoke many \syscalls, many of them can be stubbed or faked to run popular workloads.
}

\begin{figure*}
\centering
\begin{subfigure}{.495\textwidth}
  \begin{minted}[frame=single,framesep=4pt]{C}
if (getrlimit(RLIMIT_NOFILE, &limit) == -1) {
  serverLog(LL_WARNING, "Unable to obtain the current NOFILE limit,"
  "assuming 1024 & setting the max clients config accordingly.");
  server.maxclients = 1024 - CONFIG_MIN_RESERVED_FDS;
}
  \end{minted}
  \caption{Stubbing-resilient Situation (Redis).}
  \label{fig:stubbinglisting}
\end{subfigure}%
\hfill
\begin{subfigure}{.495\textwidth}
  \begin{minted}[frame=single,framesep=4pt]{C}
if (prctl(PR_SET_KEEPCAPS, 1, 0, 0, 0) == -1) {
  ngx_log_error(NGX_LOG_EMERG, cycle->log,
  ngx_errno, "prctl(PR_SET_KEEPCAPS, 1) failed");
  exit(2); /* fatal */
}
  \end{minted}
  \caption{Faking-resilient Situation (Nginx).}
  \label{fig:fakinglisting}
\end{subfigure}
\caption{
    Real world code snippets where it is effective to stub (left) and fake (right) \syscall implementations.
}
\label{fig:listing}
\end{figure*}

\paragraph{Why are Programs Resilient to Stubbing and Faking?}

Applications are able to detect and react to the failure of a \syscall.
Often, \syscall failures are non-critical and programs can take action to circumvent them.
These actions are the enabling factor of \syscall stubbing.
They include, among others (\cf \Cref{fig:heatmapdynrequired}):
\begin{itemize}
% TODO double check this
\item \textbf{Ignoring the issue.} Not all failures are consequential, and programs can simply decide to not take further action. For instance, Redis ignores when \customtt{sysinfo} \syscallno{(99)} fails to return the maximum memory size and when \customtt{ioctl} \syscallno{(16)} fails to return the resident size, as this information is only used for output to the debugging logs.
\item \textbf{Using other \syscalls}. The system call API is redundant in features: a same means can often be achieved through different \syscalls. For instance using \customtt{mmap} \syscallno{(9)} instead of \customtt{brk} \syscallno{(12)} -- a pattern from the glibc early allocator, or reallocating mappings with \customtt{mmap} \syscallno{(9)} when \customtt{mremap} \syscallno{(25)} fails, as we observe in SQLite.
\item \textbf{Falling back to safe default values}. Applications query the OS for various values to tune their behavior (max stack size and file descriptor count, processor affinity and scheduling importance, etc.). When this fails, a safe default can often be adopted. \Cref{fig:stubbinglisting} shows an example with \customtt{getrlimit} \syscallno{(97)} and \customtt{prlimit64} \syscallno{(302)} in Redis. Another example is using \customtt{ioctl} \syscallno{(16)} to query the terminal width: when this fails, Redis assumes a safe value of 80 characters.
\item \textbf{Disabling program functionalities}. Programs may also decide to simply disable the functionality that makes use of the \syscall; in certain cases, this may not even have observable consequences. For example, many applications only make use of \customtt{connect} \syscallno{(42)} through the glibc for the NSCD cache socket~\cite{NSCD_CACHE}. When \customtt{connect} fails, name caching is simply disabled.
% TODO an example of when this actually breaks a functionality would be nice to have as well
\end{itemize}

% TODO more examples of situations where faking is OK

In other cases, programs may interpret the failure conservatively and decide to abort, making stubbing impossible.
Still, in a subset of these cases, programs are overly conservative and the failure of the \syscall is in reality non-critical: if so, faking a successful return value for the \syscall, without \emph{actually} doing the work of the \syscall in question, will work.
\Cref{fig:fakinglisting} presents a concrete example in Nginx, where \customtt{prctl} \syscallno{(157)} fails to force the retaining of capabilities upon UID transition; in the context of an OS that does not have a user/kernel separation, like a unikernel, capabilities make little sense and so it is fine to fake success: faking the \syscall here will have strictly no impact on the correct execution of the software.
Similar examples are \customtt{get/setgroups} \syscallno{(115-116)}, or \customtt{setsid} \syscallno{(112)} which have, once again, no meaning in the context of a unikernel.
Still, faking OS features may also result in breaking program functionalities, \eg { }\customtt{pipe2} \syscallno{(293)} in Redis (see \S\ref{subsec:fakestub-impact}).
If the functionality is not part of the target set of application features, faking may remain a reasonable approach to achieve a first level of compatibility.

% TODO NOTE: if we had space here we could talk about exit_group (231), when we don't :)

Inversely, certain \syscalls can (almost) never be stubbed nor faked without breaking core program functionalities.
Though generalization is difficult, these system calls typically represent fundamental OS features: executing programs with \customtt{execve} \syscallno{(59)}, opening and writing to connections with \customtt{bind} \syscallno{(49)}, \customtt{listen} \syscallno{(50)}, \customtt{socket} \syscallno{(41)}, and \customtt{writev} \syscallno{(20)}, allocating memory with \customtt{mmap} \syscallno{(9)}.
We also find vectored \syscalls like \customtt{fcntl} \syscallno{(72)}, motivating our discussion in \S\ref{subsec:partial-impl}.

% TODO maybe example + discussion of situations where stubbing is NOT OK for non-functional reasons, pointing to sec 5.3

\paragraph{\SysCall Return Value Checks.}

\begin{figure}
% figure placement is tricky, adjust manually
\center
\includegraphics[width=0.9\linewidth]{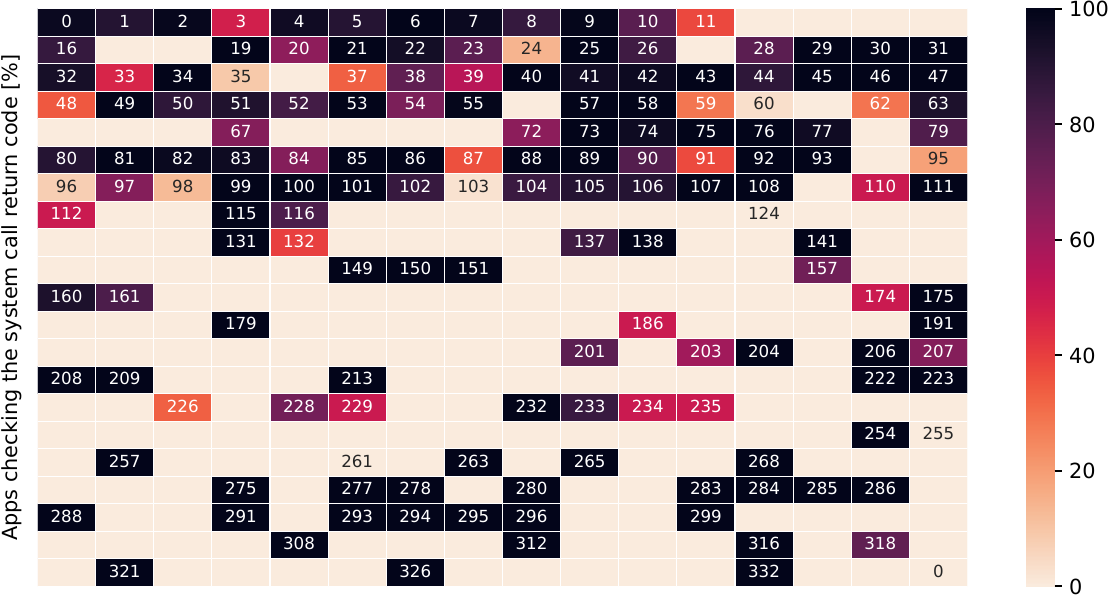}
\caption{
Apps checking \syscalls return values.
}
\label{fig:app-resilience}
\end{figure}

In addition to identifying \syscalls issued by applications, we performed a manual inspection of these applications' source code in order to gather ground truth about which \syscalls had their return values checked.
Is there a link between the presence or absence of checks, and the ability to stub or fake?
Note that we are interested here in user-written code, so we look at whether C standard library \syscalls wrappers -- not \syscalls themselves -- have their return value checked.

We choose manual inspection; building an automated static analysis method for this task is non-trivial and rather out of the scope of this paper: some programs directly check the return value, others store it in a variable which is later checked, directly or through auxiliary functions, while yet others rely on macros to do the checking.
We semi-automated the process by building scripts scanning sources for \syscall wrapper invocations and displaying their corresponding location in source files; we then manually checked this output to determine if the return code was checked or not.

Figure~\ref{fig:app-resilience} shows, for each \syscall wrapper, the number of programs that check its return value.
The majority have their return value checked.
Studying the small set of \syscalls for which no application has checks, we identify \syscalls that always succeed, \eg { }\customtt{alarm} \syscallno{(37)}, \customtt{getppid} \syscallno{(110)}, but also several that can actually fail: \customtt{getrusage} \syscallno{(98)}, \customtt{utime} \syscallno{(132)}, \customtt{inotify\_rm\_watch} \syscallno{(255)} and \customtt{futimesat} \syscallno{(261)}.
For those invoked and traced by \tool, we observe that all can be stubbed/faked for this set of applications.
Nevertheless, it would be incorrect to conclude that the ability to stub and fake is induced by the absence of checks: inversely, numerous \syscalls that are always checked can themselves often be stubbed/faked, such as \customtt{ioctl} \syscallno{(16)}, \customtt{uname} \syscallno{(63)}, or \customtt{geteuid} \syscallno{(107)}.
There is also a set of \syscalls for which only some applications feature checks.
These include \syscalls that are generally assumed to always succeed (even if they can fail) such as \customtt{clock\_gettime} \syscallno{(228)}, or freeing resources: \eg { }\customtt{close} \syscallno{(3)}, or \customtt{unlink} \syscallno{(87)}.
Generally, these can be stubbed/faked only in some applications.
Overall, we conclude that the ability to stub or fake is not a factor of the presence (or absence) of checks, but rather of the semantics of individual \syscalls and applications.

\subsection{Impact on Performance and Resource Usage}
\label{subsec:fakestub-impact}

\newcommand{\nfirst}[1]{\phantom{\textsuperscript{1}}#1\textsuperscript{1}}
\newcommand{\ntwo}[1]{\phantom{\textsuperscript{2}}#1\textsuperscript{2}}
\newcommand{\symbolempty}{$\varnothing$}
% Uncomment the line below for latexdiff
%\newcommand{\symbolempty}{-}

\begin{table*}[]
\setlength{\tabcolsep}{5pt}

\caption{Performance and resource usage (file descriptors: FD, memory usage) impact of stubbing and faking for Nginx, Redis, and iPerf3 (=\textit{App.}lications). Only systems calls with impact outside of the error margin (>3\%) in either category are displayed. ``-'' means \emph{no impact}; \textit{+X\%} means \textit{X\%} \textit{faster} or \textit{more} resource usage; \textit{-X\%} means \textit{X\%} \textit{slower} or \textit{less} resource usage.}
\label{tab:perf-ru-impact}
\center
\footnotesize
\begin{tabular}{|c|c|c|c|c|p{7.2cm}|p{2.5cm}|}
\hline
\multicolumn{1}{|c|}{\multirow{1}{*}{App.}}
            & \multirow{1}{*}{\SysCall} & Perf. Impact & FD Usage & Mem. Usage & \multicolumn{1}{c|}{\multirow{1}{*}{Explanations of Stubbing/Faking Impact}} & \multicolumn{1}{c|}{\multirow{1}{*}{Breaks\dots}} \\ \hline \hline
\multirow{4}{*}{Nginx} & \texttt{write} & +15\%  & -     & -     & Access logs are not written anymore, increasing performance.                 & \multicolumn{1}{c|}{Access Logging}                   \\ \cline{2-7}
		       & \texttt{brk}   & -      & -     & +17\% & Triggers a fallback to \texttt{mmap} in the glibc early allocator.         & \multicolumn{1}{c|}{\symbolempty}                     \\ \cline{2-7}
		       & \texttt{clone} & -      & -     & +10\% & Results in master process executing the worker loop.                         & \multicolumn{1}{c|}{Core functioning}                 \\ \cline{2-7}
	          & \texttt{sigsuspend} & -38\%  & -     & -     & Results in master process polling (busy-waiting) for events.                 & \multicolumn{1}{c|}{\symbolempty}                     \\ \hline \hline
\multirow{7}{*}{Redis} & \texttt{close} & -      & x8    & -     & FDs are not closed anymore.                                                  & \multicolumn{1}{c|}{\nfirst{\symbolempty}}            \\ \cline{2-7}
                       & \texttt{munmap}& -      & -     & +19\% & Regions are not disposed anymore.                                            & \multicolumn{1}{c|}{\ntwo{\symbolempty}}              \\ \cline{2-7}
		       & \texttt{brk}   & -      & -     & +2\%  & Triggers a fallback to \texttt{mmap} in the glibc early allocator.         & \multicolumn{1}{c|}{\symbolempty}                     \\ \cline{2-7}
 & \multirow{2}{*}{\texttt{sigprocmask}}& \multirow{2}{*}{-} & \multirow{2}{*}{-} & \multirow{2}{*}{-15\%} & Prevents creation of jemalloc background threads, resulting in memory being freed synchronously and/or at an earlier point. & \multicolumn{1}{c|}{\multirow{2}{*}{\symbolempty}} \\ \cline{2-7}
		       & \texttt{futex} & -66\%  & +94\% & -     & Inconsistent synchronization results in incorrect behavior.                  & \multicolumn{1}{c|}{Core functioning}                 \\ \cline{2-7}
		       & \texttt{pipe2} & -      & -25\% & -     & Pipes are not created anymore, resulting in less FDs.                        & \multicolumn{1}{c|}{Persistence}                      \\ \hline \hline
iPerf3                 & \texttt{brk}   & -      & -     & +11\% & Triggers a fallback to \texttt{mmap} in the glibc early allocator.         & \multicolumn{1}{c|}{\symbolempty}                     \\ \hline
\multicolumn{7}{r}{\raggedleft \textsuperscript{1}Within the maximum number of FD limits, core functioning is altered beyond this point. \textsuperscript{2}Within the limits of available memory.} \\
\end{tabular}
\end{table*}

An important concern when stubbing and faking \syscalls is whether doing so would have an effect on performance or resource usage.
Both detrimental and \emph{positive} effects are undesirable, as unintended improvements on these metrics may skew comparisons with a full-fledged baseline.
To study the question, we use \tool's ability to record performance and resource usage metrics while performing its analysis.
As described in \Cref{par:gather-nonfunctional-metrics}, \tool gathers performance metrics through user-defined scripts, and resource usage information (peak file descriptor and memory usage) through \customtt{/proc}.
For the sake of conciseness, we provide detailed results for a subset of three representative, performance-focused applications: Nginx (web server), Redis (key-value store), and iPerf3 (TCP benchmark framework).
Nginx is benchmarked with \customtt{wrk}~\cite{WRK} (HTTP requests/s), Redis with \customtt{redis-benchmark}~\cite{REDIS_BENCHMARK} (SET requests/s), and iPerf3 with an official iPerf client~\cite{IPERF} (TCP throughput).
All numbers are provided as averages of 10 runs.
Our results are visible in \Cref{tab:perf-ru-impact}.

\paragraph{Impact on Performance.}

For the majority of \syscalls, the variation in performance when stubbing or faking is within the error margin.
For the applications considered here, 3/45 \syscalls trigger a performance change when faked or stubbed.
For Nginx, stubbing/faking \customtt{write} increases performance as it prevents writing to access logs~\cite{NGINX_LOGGING} (something that test scripts do not check -- access logs are usually disabled in performance-focused settings as they are written to once per request).
It does not, however, prevent payloads from being written to, as this is done via \customtt{writev} (which, when stubbed or faked, prevents Nginx to answer requests correctly, and is detected by the test script).
Still for Nginx, stubbing or faking \customtt{rt\_sigsuspend} hurts performance, as it turns the master process' notification-based behavior into busy-waiting.
None of these alters the well-functioning of Nginx's core features as tested by the \tool test script.
Conversely, in the Redis case, faking \customtt{futex} results in synchronization issues, manifesting as a performance degradation.
This alters the core functioning of Redis, clearly indicating that faking \customtt{futex} is not a correct path to follow for compatibility, which matches intuitive expectations.
As for iPerf3, no \syscall results in performance degradation when faked or stubbed.

When such variations occur, \tool notifies the user that further investigation is needed to understand the implications (\eg on stability or scientific soundness) of stubbing or faking a particular OS feature for a given application.
This further emphasizes the need for a tool like \tool to avoid pitfalls which may cause debugging costs down the line, or skew comparisons with a full-fledged baseline.

\paragraph{Impact on Resource Usage.}

Similarly to performance, we find that faking or stubbing most \syscalls does not result in statistically significant variations in resource usage.
For the three applications considered, 4/45 \syscalls result in memory usage variations, and 3/45 in file descriptor usage variations, with one (\customtt{brk}) being caused by the libc and thus common among all three applications.

In the general case, as discussed earlier, \syscalls that allocate resources cannot be stubbed or faked: this is the case for memory allocation services such as \customtt{mmap} \syscallno{(9)}, but also for those that allocate file descriptors such as \customtt{openat} \syscallno{(257)} (see \Cref{fig:heatmapdynrequired}).
In particular cases, the claim is more nuanced; alternatives like \customtt{open} \syscallno{(2)} do not need to be implemented (\eg because \customtt{openat} is used instead, see Section~\ref{subsec:libc}).
Similarly, \customtt{brk} \emph{can} be stubbed or faked in a significant number of cases: for instance, the program exclusively uses \customtt{mmap}, and the only usage of \customtt{brk} is in the glibc initialization sequence, which is itself capable of falling back to \customtt{mmap} if \customtt{brk} does not function (at the cost of a slight memory usage increase, see \Cref{tab:perf-ru-impact}).
Another case is \customtt{pipe2}, which creates pipes at the process' demand.
Stubbing or faking it results in pipes not being created, which in turn results in an observable reduction in file descriptor count.
In the case of Redis, this breaks the persistence feature (which is often disabled in performance-focused experiments), but not the key-value store's core functionalities.

The situation is different for APIs that free resources.
In general, \customtt{munmap} and \customtt{close} can be stubbed or faked without functional impact, though resource usage will increase.
For Redis, faking or stubbing \customtt{munmap} and \customtt{close} leads to a 20\% increase in memory usage, and an 8x increase in open file descriptors under a \customtt{redis-benchmark} workload (\cf \Cref{tab:perf-ru-impact}).
Still, although these features can be stubbed or faked without sacrificing stability (as long as resources suffice), we note that the incentives to do so are lower than for other API elements; if the algorithm was developed to allocate resources, it should not be a problem to develop one that frees them.

Lastly, similarly to performance, variations in resource usage turn out to be good indicators of instability caused by stubbing or faking.
In the case of Nginx, faking \customtt{clone} results in the master process executing the worker event loop, which itself manifests as an increase in memory usage (likely because resources are left dangling).
Although functional in practice, it is not a reliable path to take for compatibility and meaningful performance comparison.
In the case of Redis, faking \customtt{futex} results in inconsistent synchronization, which itself translates into an increased number of allocated file descriptors (see \Cref{tab:perf-ru-impact}).

Beyond \syscalls that (de-)allocate resources, and those that indicate underlying instability, we identify two more classes of \syscalls which may impact resource usage (or performance):

\begin{itemize}
\item \textbf{Optimizing \syscalls:} by giving semantic indications to the kernel regarding \eg memory management policies, \syscalls such as \customtt{madvise}~\cite{MADVISE} should influence performance and resource usage.
This behavior is not visible when faking/stubbing in \Cref{tab:perf-ru-impact}: kernel hints are used rather sporadically in applications, and for those that use them (\eg Redis), the kernel did not perform actions that impacted our metrics.
Impact may be observable in other settings, \eg multi-process scenarios.
\item \textbf{System Limit Setters/Getters:} by getting/setting system defaults (\eg max stack size, number of FDs), getter/setter \syscalls like \customtt{prlimit64} (or part of \customtt{ioctl}) may also result in resource usage or performance variations.
For instance, with system defaults different from the ones in \Cref{tab:perf-ru-impact}, stubbing \customtt{prlimit64} in Redis results in 30\% lower memory usage under a \customtt{redis-benchmark} workload because the libc (stack size) and Redis (FD limits) default to values conservatively lower than the system limits.
\end{itemize}

\paragraph{Impact on Stubbing and Faking Policy}

Overall, we stress the importance of evaluating the impact of stubbing and faking on performance metrics as part of the process of deciding what to support and how.
Though most \syscalls do not impact performance metrics, some do: when the underlying reason is instability, the OS feature should never be faked; otherwise, whether or not to stub or fake should be an \emph{explicit factor} of the experimental setup and expectations on the OS prototype.
It is critical that the (positive or negative) impact of stubbing and faking must not be mistaken for that of the system's design.
Overall, we encourage authors of future systems research works to explicitly list features that they stub or fake for reproducibility and future analysis.

\insight{
Stubbing/Faking does not impact performance and resource usage in the general case.
Still, there are edge cases which may or may not indicate correctness issues.
Impact on either metric must call for special care when stubbing/faking.
}

\subsection{Partial Implementation of \SysCalls}
\label{subsec:partial-impl}

In the previous sections, we considered \syscalls as monolithic API elements.
This consideration shows its limits when investigating vectored \syscalls (\eg { }\customtt{ioctl}) or complex \syscalls like \customtt{mmap} (usable for memory allocation and file mapping, two very different purposes).
To clarify this point, we use \tool to determine the precise set of sub-\syscall features applications require.

Our insights are twofold. First, applications execute surprisingly few features from complex or vectored \syscalls.
For example, almost all applications require \customtt{arch\_prctl} \syscallno{(158)} (see \Cref{fig:heatmapdynrequired}).
However, they are far from requiring a full implementation: in fact, in all applications that we considered, this \syscall was exclusively called by the libc, which requires one single feature (\customtt{ARCH\_SET\_FS}, out of 6 in total) related to thread local storage setup.
The situation is similar for \customtt{prlimit64} \syscallno{(302)}, required by many applications: out of 16 features, only 3 are used, \customtt{RLIMIT\_CORE}, \customtt{\_NOFILE}, and \customtt{\_STACK}, the latter one being used almost exclusively as part of the libc initialization.
This is also the case for \customtt{ioctl} \syscallno{(16)}: with a benchmark load, Redis, weborf, and h2o use one single feature (\customtt{TCGETS}), Nginx two (\customtt{FIONBIO} and \customtt{FIOASYNC}), and Lighttpd none.
All of them can be stubbed.

Second, when looking at required features of \syscalls, we find that certain \syscalls such as \customtt{fcntl} typically exhibit a mix of required and fakeable/stubable features, and the required set is typically common among applications.
For instance, \customtt{F\_SETFL} is required to put file descriptors in non-blocking mode in all applications except Nginx, a critical operation for most codebases.
On the other hand, \customtt{F\_SETFD} is widely executed across applications but can always be stubbed as it is used to enable \textit{close-on-exec} on file descriptors, a non-critical operation.
In these cases, taking a look at the required \syscalls at the granularity of a \syscall would make the situation appear worse than it is in practice.

\insight{
  Several complex \syscalls do not require a full implementation to support a large number of applications.
}

% TODO ... and there are clear patterns on what can be stubbed and what cannot

\subsection{Stability of \SysCall Usage Over Time}
Once an OS prototype supports an application, how likely is it that, as the program evolves over time, additional or different \syscalls will be required, breaking the initial support?
Here we study the stability of \syscall usage by applications and libcs.

\begin{table}
\caption{
Nginx 0.3.19 \syscall usage with different glibc versions.
\Syscalls that vary because of the architecture (32/64-bit) are in \emph{italics}; other variations are in \textbf{bold}.
% we cannot sort them if we want to use the least amount of sace
}
\setlength{\tabcolsep}{4pt}
\label{tab:glibcoldnew}
{\footnotesize
\begin{tabular}{|p{0.46\linewidth}|p{0.47\linewidth}|}
\hline
\textsc{glibc 2.3.2} / 32-bit \hfill (\textit{48 \syscalls}) & \textsc{glibc 2.31} / 64-bit \hfill (\textit{51 \syscalls})\\ \hline \hline
\textit{\_llseek}, accept, access, bind, brk, clone, close,
connect, epoll\_create, \textit{fcntl64}, epoll\_ctl, epoll\_wait,
execve, exit\_group, dup2, \textit{fstat64}, \textit{geteuid32}, mkdir,
\textbf{mmap2}, \textit{setuid32}, \textbf{old\_mmap}, \textit{setgroups32}, \textbf{uname},
\textbf{open}, prctl, \textit{pread}, \textbf{pwrite}, read, rt\_sigaction,
rt\_sigprocmask, rt\_sigsuspend, \textbf{set\_thread\_area}, \textit{setgid32},
setsid, setsockopt, \textbf{recv}, socket, socketpair, \textit{stat64}, bind,
munmap, umask, getpid, getrlimit, ioctl, write, writev,
gettimeofday, listen
& read, write, close, \textit{stat}, \textit{fstat}, lstat,
\textit{lseek}, brk, rt\_sigaction, \textbf{mmap}, ioctl,
rt\_sigprocmask \textit{pread64},
setsockopt, writev, access,
sendfile, socket, munmap, accept,
connect, epoll\_wait, \textbf{mprotect},
\textbf{recvfrom}, listen, socketpair,
\textit{pwrite64}, \textit{prlimit64}, epoll\_create,
clone,execve, \textit{fcntl}, mkdir, umask,
\textit{setuid}, \textit{setgid}, \textit{geteuid}, setsid,
rt\_sigsuspend, dup2, \textit{setgroups},
\textbf{\_sysctl}, prctl, \textbf{arch\_prctl}, getpid,
\textbf{set\_tid\_address}, exit\_group,
epoll\_ctl, \textbf{openat}, \textbf{set\_robust\_list}
\\ \hline
\end{tabular}
}
\end{table}

\paragraph{Evolution: C Standard Library.}
We first study the libc, from which most \syscalls invocations generally originate.
We compiled Nginx v0.3.19 against an old version of glibc (2.3.2, from 2003) and a modern one (2.31, from 2020).
Since we were unable to run Nginx 0.3.19 with glibc 2.3.2 in 64-bit mode, we compiled and run this configuration in 32-bit.
This is likely due to these versions featuring unstable AMD64 support (the first AMD64 CPUs were released in 2003~\cite{Keltcher2003}).
The results in \Cref{tab:glibcoldnew} show that the number of used \syscalls is more or less unchanged, 48 vs 51.
Moreover, we see that most of the change in \syscall usage is caused by the deprecation of old \syscalls.
Still, there is some evolution in the types of \syscalls invoked, which we classify into two categories.
First, the recent libc uses a different version of some \syscalls due to a change of architecture (\eg it uses \customtt{pread64} instead of \customtt{pread}).
Second, the recent libc uses additional \syscalls, \eg { }\customtt{arch\_prctl} to set up TLS.
Setting aside the issue of supporting a new architecture (orthogonal to compatibility), we assume that it is the second category that would require supporting effort for updating a given compatibility layer as applications evolve.
However, we consider this effort to be low: we only count 8 new \syscalls in 17 years for this libc/application combination.
\begin{figure}
  % yes, dumb figure placement, needs manual adjustments
  \centering
  \includegraphics[width=0.9\linewidth]{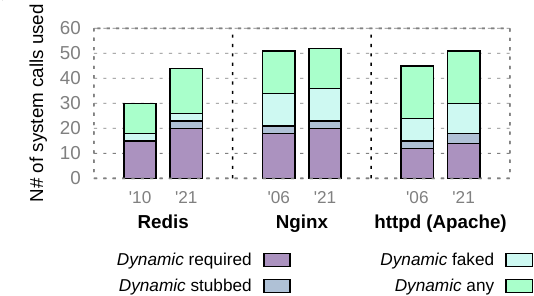}
  \caption{\Syscall usage and capacity to be stubbed/faked
      for recent (2021) and older (2005-2010) applications releases.}
  \label{fig:oldvsnew}
\end{figure}

\paragraph{Evolution: Application.}
We are now interested to see how the \syscall usage of an application changes over the years.
For this experiment, we used a modern glibc/compiler.
We explore the difference in \syscall usage of Nginx, Apache and Redis through the years and list the results in \Cref{fig:oldvsnew}.
We observe that, although the number of Linux \syscalls has increased, all applications are using roughly the same amount of \syscalls; the number of \syscalls that can be stubbed or faked also remains almost unchanged.

In all, we find the usage of \syscalls by applications and libcs to be fairly stable over time.
This is further encouragement to OS prototype developers: once you provide support for an application, you are likely to be able to keep it with minimal work for a long while.

\insight{
  Application and libc \syscall usage patterns tend to be stable over time: support is a one-time effort.
}

\subsection{C Library Impact on \SysCall Usage}
\label{subsec:libc}

Typical applications perform the majority of their \syscall invocations through the C standard library (libc).
Bypassing the libc using direct \syscall invocation happens only for functionalities rarely called by user code (\eg { }\customtt{futex}) or newer \syscalls for which libcs do not provide a wrapper: we counted around 51 \syscalls (58 including removed/unimplemented \syscalls) that do not have a wrapper as of glibc 2.33.
In this case, applications wishing to invoke them use the \customtt{syscall} function.
Setting aside these special cases, we find that the libc implementation greatly influences the \syscall API usage of applications.
This is due to two main factors: (1) the libc initialization sequence and (2) the choice of \syscall alternatives (\eg { }\customtt{openat} \vs { }\customtt{open}).

\paragraph{Libc Initialization Sequence.}
The initialization sequence is the libc code executed from the program entry point until the user's \customtt{main} function is invoked.
The \syscalls invoked by that code will be by construction present in any binary linked against that libc and constitute the minimum set of \syscalls an OS should implement to support this libc.
To study initialization sequences, we recorded the \syscall usage of a trivial application printing "Hello, world!" across two libcs, glibc (version 2.28) and musl (version 1.2.2), for a dynamically- and a statically-linked executable.
Results in Table~\ref{tab:libchelloworld} show that the number and types of \syscalls executed vary: glibc's initialization sequence invokes for dynamically compiled binaries 2.5x more \syscalls \vs musl, and 1.8x more for statically compiled programs.
The \syscalls invoked also change: glibc is not a strict superset of musl and out of 18 \syscalls in total, only 6 are common to both libcs for dynamic, 3 for static (and 3 overall).

\paragraph{\SysCall Alternatives.}
Some discrepancies are due to the libcs choosing different \syscall alternatives to perform the same task.
For example, glibc uses \customtt{write} for \customtt{printf}, \vs { }\customtt{writev} for musl.
Similarly, musl uses \customtt{ioctl} to check that the TTY is writable, while glibc uses \customtt{fstat}.
Finally, glibc uses \customtt{openat}, \customtt{read}, \customtt{mmap}, and \customtt{mprotect} to map the libc into the address space, an operation that musl achieves by embedding the libc into the linker itself, avoiding these \syscalls entirely.
Other differences are caused by libc-specific initialization and debugging features.
For example, even in single-threaded programs, musl will call \customtt{set\_tid\_address} during TLS initialization, something that glibc does not.
Glibc, on the other hand, uses \customtt{uname} to ensure that the kernel is recent enough, \customtt{readlink} to expand \customtt{\$ORIGIN} with statically compiled binaries, and \customtt{access} for a debugging feature; none of these used by musl's initialization sequence.

\begin{table}[]
\caption{
\Syscall API usage of a hello world application across glibc (2.28) and musl (1.2.2).
Apart from \customtt{exit\_group} and
\customtt{write}/\customtt{writev}, this set corresponds to the libc
initialization sequence.
Differing \syscalls are in bold.
}
\setlength{\tabcolsep}{4pt}
\label{tab:libchelloworld}
{\footnotesize
\begin{tabular}{|p{0.5\linewidth}|p{0.42\linewidth}|}
\hline
\underline{glibc} & \underline{musl} \\ \hline
\textit{28 \syscalls (dynamic binary)} & \textit{11 \syscalls (dynamic binary)} \\ \hline
execve {\scriptsize(1x)}, brk {\scriptsize(3x)}, arch\_prctl {\scriptsize(1x)}, exit\_group {\scriptsize(1x)}, \textbf{access} {\scriptsize(1x)}, \textbf{openat} {\scriptsize(2x)}, \textbf{fstat} {\scriptsize(3x)}, mmap {\scriptsize(7x)}, \textbf{close} {\scriptsize(2x)}, \textbf{read} {\scriptsize(1x)}, mprotect {\scriptsize(4x)}, \textbf{munmap} {\scriptsize(1x)}, \textbf{write} {\scriptsize(1x)} & execve {\scriptsize(1x)}, brk {\scriptsize(2x)}, arch\_prctl {\scriptsize(1x)}, exit\_group {\scriptsize(1x)}, \textbf{writev} {\scriptsize(1x)}, mmap {\scriptsize(1x)}, mprotect {\scriptsize(2x)}, \textbf{ioctl} {\scriptsize(1x)}, \textbf{set\_tid\_address} {\scriptsize(1x)} \\ \hline
\textit{11 \syscalls (static binary)} & \textit{6 \syscalls (static binary)} \\ \hline
execve {\scriptsize(1x)}, arch\_prctl {\scriptsize(1x)}, exit\_group {\scriptsize(1x)}, \textbf{brk} {\scriptsize(4x)}, \textbf{fstat} {\scriptsize(1x)}, \textbf{write} {\scriptsize(1x)}, \textbf{uname} {\scriptsize(1x)}, \textbf{readlink} {\scriptsize(1x)} & execve {\scriptsize(1x)}, arch\_prctl {\scriptsize(1x)}, exit\_group {\scriptsize(1x)}, \textbf{writev} {\scriptsize(1x)}, \textbf{ioctl} {\scriptsize(1x)}, \textbf{set\_tid\_address} {\scriptsize(1x)} \\ \hline
\end{tabular}
}
\end{table}

\insight{
  The \emph{choice of libc} and \emph{linking type} strongly influences \syscall usage: as much as 4.5x fewer \syscalls between dynamic glibc and static musl.
}

\section{Discussion: Pitfalls \& Future Works}
\label{sec:discussion}

As discussed throughout this work, there are pitfalls to developing OS compatibility layers with dynamic analysis, stubbing, faking, and partial support techniques.

\paragraph{Impact on Stability.}
Dynamic analysis, stubbing, faking, and partial support techniques, bring the concern of stability: \emph{do we trade off correctness to reduce porting time?}
\tool assumes that users are able to evaluate the functionality of application features they aim to support by specifying a set of tests (\S\ref{subsec:testapps}).
The tool ensures that this set of tests can be passed reliably, over multiple runs, when applying stubbing, faking, and partial support techniques.
\tool can also ensure that performance, resource usage, and any other metric, remains stable (\S\ref{subsec:fakestub-impact}).
Under this assumption, stability issues outside users' target feature range are not in the problem scope of \tool, or our study.
Still, perfect correctness cannot be guaranteed, and compatibility bugs may hide in incomplete or buggy tests, varying test environments, etc. We believe that these are reasonable trade-offs to be made in transitional development stages of a new OS.

\paragraph{Impact on Evaluation Metrics.}
Assuming stability, another concern remains: \emph{do we trade off (or simply influence) performance, resource usage, or any other metric for porting time?}
This is most relevant as early OS prototypes must be able to compare, in a sound manner, properties with full-fledged baseline OSes.
We show that, although the majority of \syscalls do not influence performance metrics when stubbed, faked, or partially supported, there \emph{are} pitfalls: even when reliably passing tests, these techniques can result in visible performance or resource usage variations (\S\ref{subsec:fakestub-impact}).
\tool improves on the state of the art, which does not consider this problem, by evaluating these costs systematically and early, to provide strong evidence that achieved support does not impact chosen metrics.
Still, it remains impossible to formally guarantee that these metrics will be unaffected in all cases.
We believe that this too constitutes a reasonable trade-off in development stages.

\medskip
Overall, dynamic analysis, stubbing, faking, and partial support should not be seen as end-goals for production-ready compatibility, but as a transitional, ``necessary evil'' in development phases.
The takeaway of this paper should not be that most of the \syscall API is irrelevant, or that static analysis is impertinent in engineering compatibility layers; each corresponds to distinct life cycle phases in the development of new OS.
As we show, static analysis is not appropriate in earlier stages, however its output should decisively be a target in later stages of development, and full support should eventually come to achieve high levels of correctness assurance.

Looking forward, we plan to improve \tool with support for other analysis metrics, such as identifying standard application-specific logs and error message formats, or network and file system usage statistics, to better detect silent faults and effects of stubbing, faking, and partial support techniques.
We believe that there remain many interesting research opportunities in application analysis for compatibility that should be explored in future works.
Future research avenues include exploring speeding up the analysis by transferring knowledge across applications, and generally using machine learning techniques to identify patterns over the data set, at scale, and generating application-specific workloads.

\section{Related Work}
\label{sec:related-works}

\paragraph{OS Compatibility Layers.}
Many research and prototype OSes have implemented compatibility layers to transparently support legacy software.
An early example~\cite{K42_LINUX} presents a compatibility layer for Linux applications implemented in the K42~\cite{K42} OS.
Similarly to our work, the authors note that to be widely adopted, an OS must provide good support for existing applications, and that emulating the Linux API is the best way to achieve this goal without requiring modification of target applications.
In another study~\cite{POSIX_PICOPROCESS}, researchers propose a POSIX compatibility layer for the Embassies~\cite{EMBASSIES} system.
This work presents the construction of the compatibility layer, which is realized in a fully ad-hoc way.
As we demonstrate, this process can be highly optimized with \tool.
Still, the authors make some observations similar to ours, in particular the fact that some \syscalls are ``failure-oblivious'' (\ie they can be stubbed) and others are ``neutered'' (they can be faked).
Other works proposed compatibility layers for new monolithic~\cite{LINUXULATOR, KERLA, ZEPHYR}, libOS~\cite{GRAPHENE, GRAPHENE_SGX, UNIKRAFT, OSV, HERMITUX, LUPINE, GVISOR} or micro-kernels~\cite{FUCHSIA, REACTOS}, web browsers~\cite{BROWSIX}, for running applications within the Linux kernel~\cite{NOLIBC}, as well as various OS interoperability layers for existing kernels~\cite{WSL, WINE, PROTON, LINUXULATOR, DOCKER_ON_ARM}.
To the best of our knowledge, all these compatibility layers have been developed in an organic way.

\paragraph{Libc-Based Compatibility Layers.}
Some works~\cite{NEWLIB, OSV} approach compatibility at the libc level, instead of the \syscall API.
Though most \syscalls are performed through the libc, prior works have shown that interfacing at the libc level leads to weaker degrees of compatibility~\cite{HERMITUX_TC} because many programs do issue system calls outside the libc (500+ ELF Debian 10 executables fall into that category~\cite{HERMITUX_TC}).
Thus, we focus on compatibility at the \syscall level.

\paragraph{Linux \& POSIX APIs Studies.}
Past work studied the usage of the Linux~\cite{LINUX_API} and POSIX~\cite{POSIX_API} APIs by applications.
Tsai et al.~\cite{LINUX_API} use binary static analysis to measure the \syscalls and pseudo files required by a large set of binaries from the Ubuntu 15.04 archive.
Even for the most minimal Ubuntu installation, the study reports that 224 \syscalls, 208 \customtt{ioctl}/\customtt{prctl}/\customtt{fcntl} codes and 100+ pseudo files require support.
Our results demonstrate that static binary analysis is overly pessimistic.
Using dynamic analysis, \tool shows that the amount of OS features required to run standard benchmarks or even full test suites is actually much lower.

Another study~\cite{POSIX_API} leverages both static and dynamic analysis to measure applications' POSIX API usage.
Unlike this work, the authors' goal is not to determine and optimize compatibility efforts, but to study the evolution of the POSIX interface and identify emerging/missing abstractions.
Though the study provides valuable insights for building a compatibility layer at the POSIX (\ie libc) level~\cite{OSV, LUPINE}, past studies showed that the Linux API (mainly \syscalls) provided a higher degree of compatibility~\cite{HERMITUX_TC}: the authors themselves~\cite{POSIX_API} note that many applications (\eg Go apps) circumvent POSIX to use OS specific APIs.

\section{Conclusion}
\label{sec:conclusion}
We propose \tool, an efficient method to determine and prioritize OS features new compatibility layers should implement to provide support for as many applications as possible, as early as possible.
Applying \tool to 100+ applications, we provide examples of support plans, demonstrate high engineering effort savings, and study our measurements in depth.
A significant number of \syscalls identified as needed by previous works are actually not required for those applications to run.
These results bring a message of hope to the level of compatibility a new OS must provide in order to support mainstream applications, and should provide encouragement to ongoing and future research OS development efforts.

\section*{Acknowledgments}

We thank the anonymous reviewers and our shepherd, Donald E. Porter, for their
insights. This work was funded by a studentship from NEC Labs Europe, a
Microsoft Research PhD Fellowship, UK’s EPSRC grants \grantno{EP/V012134/1}
(UniFaaS), \grantno{EP/V000225/1} (SCorCH), and the EPSRC/Innovate UK grant
\grantno{EP/X015610/1} (FlexCap), as well as EU H2020 grants \grantno{825377}
(UNICORE), \grantno{871793} (ACCORDION) and \grantno{758815} (CORNET). UPB
authors were supported by VMWare gift funding.

\bibliographystyle{plain}
\bibliography{bib}

\end{document}